\documentclass[11p,a4paper]{article}
\pdfoutput=1
\usepackage{graphicx}
\usepackage{jheppub}
\usepackage[utf8]{inputenc}
\usepackage{latexsym,amsfonts,amsmath,amscd,amssymb,epsf}
\usepackage{hyperref}
\usepackage{soul,xcolor}
\usepackage{natbib}
\usepackage{enumitem}
\usepackage{graphicx}
\usepackage{subfig}
\usepackage{url}

\newcommand{\beq}{\begin{equation}}
\newcommand{\eeq}{\end{equation}}

\DeclareMathAlphabet{\pazocal}{OMS}{zplm}{m}{n}

\renewcommand{\v}[1]{\mathbf{#1}}
\usepackage{bm}
\let\v=\bm

\title{Towards an Interpretation of the First Measurements of Energy Correlators in the Quark-Gluon Plasma}

\author[a]{Carlota Andres,}
\author[b]{Fabio Dominguez,}
\author[c]{Jack Holguin,}
\author[d]{Cyrille Marquet,}
\author[e]{Ian Moult}
\emailAdd{carlota@lip.pt}
\emailAdd{fabio.dominguez@usc.es}
\emailAdd{jack.holguin@manchester.ac.uk}
\emailAdd{cyrille.marquet@polytechnique.edu} 
\emailAdd{ian.moult@yale.edu} 

\affiliation[a]{Laborat\'orio de Instrumenta\c{c}\"ao e F\'isica Experimental de Part\'iculas (LIP), Av.~Prof.~Gama Pinto, 2, 1649-003 Lisbon, Portugal}

\affiliation[b]{Instituto Galego de F\'isica de Altas Enerx\'ias (IGFAE), Universidade de Santiago de Compostela, Santiago de Compostela 15782, Spain}
\affiliation[c]{Consortium for Fundamental Physics, School of Physics \& Astronomy, \\
	University of Manchester, Manchester M13 9PL, United Kingdom}
 \affiliation[d]{CPHT, CNRS, \'Ecole polytechnique,
Institut Polytechnique de Paris, 91120 Palaiseau, France}
\affiliation[e]{Department of Physics, Yale University, New Haven, CT 06511}
\newcommand{\comment}[1]{}

\newcommand{\td}{\mathrm{d}}

\renewcommand{\>}{\right\rangle}
\newcommand{\<}{\left\langle}

\newcommand{\qqqquad}{\qquad\qquad\qquad}

\newcommand{\As}{\alpha_{\mathrm{s}}}

\newcommand{\cE}{\mathcal{E}}

\newcommand{\Sec}[1]{section~\ref{#1}}

\begin{document}

\abstract{Energy correlators have recently been proposed as a class of jet substructure observables that directly link experimental measurements of the asymptotic energy flux with the field theoretic description of the underlying microscopic dynamics.
This link holds particular promise in heavy-ion physics, where both experimental measurements and theoretical interpretations are inherently complex.
With recent measurements of energy correlators in proton-proton collisions, the first measurement of these observables on inclusive jets in heavy-ion collisions underscores the importance of a theoretical understanding of their behavior in this complex environment.
In this manuscript, we extend our previous calculations to account for several effects necessary for a qualitative understanding of the behavior of energy correlators on inclusive jets in heavy-ion collisions. Through a semi-analytic approach implemented in a hydrodynamically expanding quark-gluon plasma (QGP), we account for medium-induced radiation with leading broadening effects, selection biases arising from energy loss, and a description of the confinement transition.
Our results represent a crucial first step towards interpreting the measurements of energy correlators on inclusive jets in heavy-ion collisions, which marks a significant milestone in connecting heavy-ion experiment and fundamental quantum field theory, in the quest to disentangle the microscopic dynamics of the QGP.
}

\maketitle
%%%%%%%%%%%%%%%%%%%%%%%%%%%%%%%%%%%%%%%%%%%%%%%%%

%%%%%%%%%%%%%%%%%%%%%%%%%%%%%%%%%%%%%%%%%%%%%%%%%
\section{Introduction}
\label{sec:intro}
%%%%%%%%%%%%%%%%%%%%%%%%%%%%%%%%%%%%%%%%%%%%%%%%%

The ability to create the quark-gluon plasma (QGP) in collider experiments \cite{Gyulassy:2004zy,PHOBOS:2004zne,Muller:2006ee,Muller:2012zq} offers an unparalleled opportunity to study quantum chromodynamics (QCD) under extreme conditions. For recent reviews, see, for instance, \cite{Busza:2018rrf,Cao:2020wlm,Arslandok:2023utm,CMS:2024krd,ALICE:2022wpn}. In heavy-ion collisions, concurrently with the formation of a hydrodynamically expanding QGP, high-energy jets are produced in the underlying hard scattering process, providing a unique window into the QGP microscopic dynamics \cite{Apolinario:2022vzg, Cunqueiro:2021wls,Connors:2017ptx}. However, the intricate nature of the QCD parton shower, the subsequent hadronization process, and the large underlying event complicate the interpretation of the jet measurements in heavy-ion collisions, making it exceedingly challenging to extract information regarding the microscopic interactions of quarks and gluons within the QGP.

In recent years, significant progress has been made on both the experimental and theoretical fronts in understanding  the inner structure of jets, a field commonly referred to as jet substructure  \cite{Larkoski:2017jix,Asquith:2018igt,Marzani:2019hun}. Novel theoretical developments have reshaped the investigation of jet substructure by framing it as the characterization of a quantum mechanical system through the asymptotic energy flux. These techniques are particularly powerful when dealing with large numbers of particles, enabling a shift from studying exclusive amplitudes to describing the system in terms of the statistical properties of the flux. This approach is reminiscent of methodologies in cosmology \cite{Maldacena:2002vr,Arkani-Hamed:2015bza,Arkani-Hamed:2018kmz}.

This reformulation has been possible through the introduction of energy correlators \cite{Basham:1977iq,Basham:1978bw,Basham:1978zq,Basham:1979gh,Hofman:2008ar} in the study of jet substructure \cite{Dixon:2019uzg,Chen:2020vvp,Komiske:2022enw}. These observables establish a direct link between experimental data and the underlying quantum field theory, bypassing the need for complex algorithms, such as declustering.  Due to their simplicity, energy correlators hold significant promise for jet substructure studies, particularly in the complex heavy-ion environment, where they offer great potential to unveil QGP microscopic dynamics \cite{Andres:2022ovj, Andres:2023xwr,Andres:2023ymw}. While these advancements, alongside other recent approaches \cite{Yang:2023dwc, Barata:2023zqg, Barata:2023bhh}, mark initial steps toward understanding medium modifications of energy correlators, they are constrained by reliance on simplified models for the medium and/or jet-medium interactions. Therefore, there is an urgent need to deepen this understanding and identify which observed features might be model-dependent. 

Furthermore, recent measurements of energy correlators in proton-proton collisions \cite{talk3, Tamis:2023guc, talk2,talk4,CMS:2024mlf,talk6} where scaling laws in the asymptotic energy flux have been directly verified, and the confinement transition has been observed, have prompted efforts to measure energy correlators in inclusive jets in heavy-ion collisions. As these measurements approach, it becomes even more crucial to move beyond these initial steps \cite{Andres:2022ovj, Andres:2023xwr,Andres:2023ymw,Yang:2023dwc, Barata:2023zqg, Barata:2023bhh} and understand how the complexities of real-world heavy-ion collisions, such as selection biases in inclusive measurements or the presence of a dynamical medium, manifest in energy correlators. This understanding is essential for accurately interpreting heavy-ion energy correlator inclusive measurements and effectively exploiting the potential of these observables to unravel the microscopic dynamics of the QGP.

In this manuscript, we expand upon the calculations presented in \cite{Andres:2022ovj,Andres:2023xwr,Andres:2023ymw} to advance the theoretical description of these observables. We incorporate a variety of additional physical effects that were previously overlooked,  but are crucial for a more realistic description of energy correlators in heavy-ion inclusive jet measurements. Specifically, using a semi-analytic approach, we now account for the effects of a hydrodynamically expanding medium,  a description of the confinement transition, and selection biases arising from energy loss. We further go beyond the semi-hard approximation to the in-medium splittings and concurrent multiple scatterings used in \cite{Andres:2022ovj,Andres:2023xwr,Andres:2023ymw}, in order to account for corrections due to transverse momentum broadening. Our primary objective is to provide an analytical framework that yields clear physical insights into each of these effects, which will be key for interpreting the first energy correlator measurements in heavy-ion collisions. This work represents a significant advancement over \cite{Andres:2022ovj, Andres:2023xwr, Andres:2023ymw}. However, it leaves open many areas for future improvement, in particular the incorporation of medium response \cite{Cao:2020wlm} and the further improvement of perturbative calculations \cite{forwardcite}.

An outline of this paper is as follows. In \Sec{sec:eecs} we present a brief overview of energy correlator observables, and our approach to compute their modifications in the medium. In \Sec{sec:inclusive} we detail the calculation the two-point energy correlator in heavy-ion inclusive jet samples including: in-medium splittings at leading order in $\alpha_s$ within either a single scattering or multiple scattering approach with leading broadening effects; a dynamically expanding medium; energy loss; and a model of the non-perturbative transition. In \Sec{sec:results} we present our numerical results and highlight the role of the selection bias due to energy loss on the energy correlator distribution of inclusive heavy-ion jets. We conclude in \Sec{sec:conc}.

\vspace{0.25cm}
{{\bf Note:} The research presented in this paper was motivated by the goal of understanding the first measurement of the energy correlators in heavy-ion collisions by the CMS experiment. All calculations were performed while the data was blinded. The CMS measurement was first presented by Jussi Viinikainen at the ``Energy Correlators at the Collider Frontier" workshop on July 9, on the same day as the release of this paper. Details of the measurement can be found at} \url{https://indico.mitp.uni-mainz.de/event/358/timetable/#20240712.detailed}, and \cite{CMS-PAS-HIN-23-004}.

%%%%%%%%%%%%%%%%%%%%%%%%%%%%%%%%%%%%%%%%%%%%%%%%%
\section{Energy Correlators}
\label{sec:eecs}
%%%%%%%%%%%%%%%%%%%%%%%%%%%%%%%%%%%%%%%%%%%%%%%%%

In this section we provide a review of the basic properties of energy correlator observables and their modifications in a QCD medium. We start by outlining in \Sec{sec:review_EEC} the properties of the energy correlator observables in (vacuum) QCD, focusing on their operator definitions and the relationship between their behavior across angular scales and the underlying field theory. Subsequently, in \Sec{sec:EEC_HI} we focus on how  energy correlators are modified in heavy-ion collisions. 

%%%%%%%%%%%%%%%%%%%%%%%%%%%%%%%%%%%%%%%%%%%%%%%%%
\subsection{Energy Correlator Observables}
\label{sec:review_EEC}
%%%%%%%%%%%%%%%%%%%%%%%%%%%%%%%%%%%%%%%%%%%%%%%%%

Energy correlator observables are formulated as correlation functions, $\langle \Psi | \cE(\vec n_1) \cE(\vec n_2)\cdots \cE(\vec n_k) |\Psi \rangle$ of the energy flow operator \cite{Sveshnikov:1995vi,Tkachov:1995kk,Korchemsky:1999kt,Bauer:2008dt,Hofman:2008ar,Belitsky:2013xxa,Belitsky:2013bja,Kravchuk:2018htv},
\begin{align}
    \cE(\vec n) = 
    \lim_{r\rightarrow \infty} \int \mathrm{d}t \,r^2 n^i \,
    T_{0i}(t,r\vec{n})\,,
    \label{eq:energy_flux}
\end{align}
in some state $|\Psi \rangle$, which we will ultimately take to be a jet in a heavy-ion collision. The linear weighting in energy ensures infrared and collinear safety of these observables, while also suppressing soft radiation.  Although the initial interest in these observables stemmed from studies in conformal field theory, their significance goes beyond conformal properties.  Rather, their breakthrough lies in their ability to establish a direct link between correlation functions of a fundamental operator within the underlying theory and weighted cross-sections measured in collider experiments. This is underscored by the fact that is has been demonstrated that energy correlators observed in collider processes can be computed as \cite{Sterman:1975xv,Hofman:2008ar}
\begin{align}
    \langle \Psi | \cE(\vec n_1) \cdots \cE(\vec n_k) |\Psi \rangle = \sum_{h_1 \cdots h_k} \int \td \sigma_{h_1 \cdots h_k} ~ E_{h_1} \cdots E_{h_k} \delta^{(2)}(\vec n_{h_1}-\vec n_1) \cdots \delta^{(2)}(\vec n_{h_k}-\vec n_k), \label{eq:EECcrosssectiondef}
\end{align}
where $\td \sigma_{h_1 \cdots h_k}$ is the inclusive cross-section for producing $k$ hadrons with energies $E_{h_1} \cdots E_{h_k}$ at directions $\vec n_{h_1} \cdots \vec n_{h_k}$. Eq.~\eqref{eq:EECcrosssectiondef} remains valid even if we replace the set of hadrons $\{h_1 \cdots h_k\}$ with a set of $k$ idealized detector cells situated at directions $\vec n_{h_1} \cdots \vec n_{h_k}$, which measure energy deposits $E_{h_1} \cdots E_{h_k}$, which clearly shows that the bridge provided by energy correlators between measurements obtained from detectors and the underlying theory eliminates the need to reconstruct individual particle states.  Other compelling aspects driving theoretical interest in energy correlator observables is their lack of reliance on algorithms rooted in a classical and perturbative view of the jet formation, such as declustering. This property renders them particularly well-suited for the complex heavy-ion environment, where a combination of perturbative and non-perturbative physics prevails. It is important to note that, for practical jet substructure studies, energy correlators must be measured inside a jet, which is identified using a robust clustering algorithm. Clustering is an infra-red and collinear safe process which only modifies structures at the highest energy scales within a process, where asymptotic freedom guarantees a parton-like description. Therefore, away from a jet boundary, energy correlations can be reliability studied within a clustering prescription.

An additional experimental advantage of formulating jet substructure in terms of energy correlators is the ability to systematically incorporate the use of tracking information into theoretical calculations. There has been significant recent progress in this direction \cite{Chang:2013rca,Chang:2013iba,Jaarsma:2023ell,Chen:2022pdu,Chen:2022muj,Jaarsma:2022kdd,Li:2021zcf}, which has enabled high order perturbative calculations and resummation of energy correlators on tracks. The initial measurement of the energy correlators in a heavy-ion experiment is performed on tracks, and we believe that this will be the standard going forward, and will enable numerous new measurements. 

In this paper, our focus is on the simplest case of the two-point correlator $\langle \Psi | \cE(\vec n_1) \cE(\vec n_2) |\Psi \rangle$, computed solely as a function of the angle between the two detectors. In the case of hadrons colliders, where longitudinal-boost invariant coordinates are used, the angle between the detectors is often labeled $R_{L}$. Specifically, we study the distribution:
\beq
    \frac{\td \Sigma}{\td R_L} = \int \td \vec n_{1}\, \td \vec n_{2}\, \frac{\< \cE(\vec n_1) \cE(\vec n_2) \>}{Q^{2}} \,\delta(\Delta R_{12} - R_L)\,, 
    \label{eq:distribution}
\eeq
where $Q$ is an appropriate hard scale\footnote{In proton-proton collisions, for jets produced at central rapidities  $Q$ can be set  to the reconstructed jet energy.}, and $\Delta R_{12}$ denotes the boost invariant angle between particles $1$ and $2$. We will often refer to this observable as EEC. 

In a conformal theory, the behavior of $\td \Sigma/\td R_L$ follows a power law with respect to $R_L$, where the exponent is determined by twist-two operators \cite{Hofman:2008ar,Kologlu:2019mfz} that emerge as the leading contributions from the operator product expansion in the $R_L\to 0$ limit. In the context of high energy proton-proton collisions, this power-law persists, but its exponent is slightly modified  by single logarithmic corrections arising from the resummation of the QCD running coupling \cite{Chen:2021gdk,Chen:2022jhb}.  We note that this scaling behavior is known to very high accuracy in QCD (for instance, for \eqref{eq:distribution} at NLO+NNLL see \cite{Dixon:2018qgp}), providing us with a well-controlled baseline against which to identify heavy-ion modifications. However, since in QCD the asymptotic states are hadrons, the confinement transition must break the perturbative power-law behavior of energy correlators. This disruption occurs abruptly at an angular scale $\theta_{\rm NP} \sim \Lambda_{\text{QCD}}/Q$ below which the correlator transitions to a different scaling law dictated by non-interacting hadrons. Recent  measurements of energy correlators in  jets proton-proton (p-p) collisions at  RHIC \cite{talk3,Tamis:2023guc} and at the LHC \cite{,talk2,talk4,CMS:2024mlf,talk6} have successfully observed both the perturbative scaling law and the transition to the non-perturbative regime. The former has already been proven instrumental in extracting the strong coupling constant $\alpha_s$ from jet substructure studies~\cite{CMS:2024mlf,talk4}, and the latter may shed light on the QCD confinement dynamics \cite{Lee:2023npz}.

Similar to how non-perturbative dynamics sharply breaks the conformal behavior at $R_L \sim \theta_{\rm NP}$ in proton-proton collisions, the presence of additional scales due to QGP formation in heavy-ion collisions is expected to result in distinct changes in the energy correlator behavior. These modifications, which cannot be attributed to other factors such as variations in quark/gluon fractions,  open the door to unraveling the microscopic dynamics of the QGP using energy correlators.

\subsection{Modification of Energy Correlators in Heavy-Ion Collisions}
\label{sec:EEC_HI}

In heavy-ion collisions, energy loss is often expected to be the primary source of medium modifications to jet observables. Energy loss contributes to an observable's spectrum through soft Sudakov-logarithms, which are proportional to the color structures present throughout the jet's evolution. Consequently, an accurate account of the jet's history in the medium is crucial to interpret classical jet-substructure observables. In contrast, the single logarithmic nature of energy correlators prevents them from being swamped by Sudakov physics. Instead, EEC observables within jets are sensitive to perturbative collinear physics in the $ 1 \gg R_{L} \gg \theta_{\rm NP}$ regime, providing us with a fundamentally different probe of the QGP. As a result, energy correlators hold the potential to be sensitive to other phenomena, such as hard in-medium splittings or the medium backreaction~\cite{Cao:2020wlm}. 

This perspective guided the first analytical studies of the behavior of the large-angle ($1 \gg R_{L} \gg \theta_{\rm NP}$) regime of energy correlators in heavy-ion jets \cite{Andres:2022ovj,Andres:2023xwr,Andres:2023ymw,Barata:2023bhh}. As first pointed out in \cite{Andres:2022ovj}, medium-induced splittings induce local changes in the shape of the energy correlator spectra at  angles in this region. These modifications break the otherwise power-law scaling of the observable, and thus cannot be mimicked by modifying the quark/gluon fraction. In \cite{Andres:2022ovj, Andres:2023xwr, Andres:2023ymw, Barata:2023bhh}, these medium-induced splittings were incorporated to the EEC \eqref{eq:distribution}  at leading order (LO) in $\alpha_s$   using different jet formalisms. All these approaches rely on simplifying approximations. For instance, the multiple scattering approach used in \cite{Andres:2022ovj,Andres:2023xwr} neglects any momentum broadening, while \cite{Barata:2023bhh} incorporates a significantly less restrictive framework that includes broadening but is only available for $\gamma \to q\bar q$ splittings and for a very limited region of the parameter space \cite{Isaksen:2023nlr}. Additionally, all these analytical approaches were implemented within a static ``brick'' description of the QGP. One of the main objectives of this paper is to improve the perturbative calculation of the LO medium-induced contributions to the EEC by relaxing some of the previously employed approximations for the in-medium splittings while accounting for the QGP dynamical evolution. This is the focus of section~\ref{subsec:HO}.

Contributions from radiative energy loss to the substructure within a jet are higher-order than those from medium-induced splittings when the $R_{L} \gg \theta_{\rm NP}$ regime of the EEC \eqref{eq:distribution} is computed, as discussed in detail in \cite{ Andres:2023xwr}. However, it is important to note that any distinctive feature of the EEC spectra appears at angles fixed by the ratio of the relevant dynamical scale against the hard scale $Q$. Hence, on heavy-ion inclusive jets measurements, they will be influenced by the selection bias relative to p-p due to energy loss. Since jets are typically selected based on their reconstructed energy, and heavy-ion jets lose energy in the QGP, comparisons of inclusive jet samples in p-p and heavy-ion collisions at the same reconstructed energy do not correspond to the same jet population, as they were produced at systematically different hard scales.\footnote{Selection biases in heavy-ion jets  have been extensively studied. See, for instance, \cite{Baier:2001yt,dEnterria:2009xfs,Renk:2012ve,Milhano:2015mng,Spousta:2015fca, Casalderrey-Solana:2015vaa,Casalderrey-Solana:2016jvj,Caucal:2018dla,Casalderrey-Solana:2018wrw, Brewer:2018dfs,Casalderrey-Solana:2019ubu,Caucal:2020xad,Du:2020pmp,Apolinario:2021olp,Brewer:2021hmh}.}  In the EEC spectrum, the transition to hadronization occurring at angular scales around $R_L \sim \theta_{\rm NP}$  will be influenced by these selection bias effects. Even if the hadronization dynamics is universal between p-p and heavy-ion collisions, this non-perturbative transition will appear modified in heavy-ion inclusive jets compared to p-p jets due to perturbative effects of energy loss. This highlights the importance of disentangling this selection bias effect for identifying genuine QGP modifications in energy correlator observables. To qualitatively understand how this selection bias modifies the hadronization transition  in heavy-ion inclusive jets, we work from the starting point that the measured jet energy scale, $\tilde{Q}$, is an extensive property of a jet and it is not a substructure measurement. Importantly, and unlike traditional Sudakov-sensitive approaches to jet substructure, this means that the qualitative effects of the selection bias can be disentangled whilst treating a jet as a single coherent radiating object (see section~\ref{sec:weights}). To understand the selection bias effects on the whole EEC spectrum, we introduce a simple model for the hadronization transition in section~\ref{sec:transition}.

The main outcome of this work is to incorporate the aforementioned effects to obtain an analytic calculation that can meaningfully provide a qualitative description of the features of the EEC in heavy-ion inclusive jets. We believe that for interpreting initial measurements, this theoretically controlled qualitative understanding is crucial. The following section details how these effects were implemented in our formalism.

%%%%%%%%%%%%%%%%%%%%%%%%%%%%%%%%%%%%%%%%%%%%%%%%%
\section{Towards an Understanding of Energy Correlators on Inclusive Jets}
\label{sec:inclusive}
%%%%%%%%%%%%%%%%%%%%%%%%%%%%%%%%%%%%%%%%%%%%%%%%%

In this section we extend our calculations from  \cite{Andres:2023ymw,Andres:2023xwr,Andres:2022ovj} to incorporate several effects necessary for the description of the energy correlators in inclusive heavy-ion jet samples. Namely, we include broadening corrections to the medium-induced formalism with multiple scattering, the impact of a dynamically evolving medium, energy loss, and a description of the confinement transition. Our goal is not to provide quantitative predictions of the two-point energy correlator to be measured in heavy-ion inclusive jets, but rather to offer significant insights into the effects that these physical mechanisms have on this observable. Additionally, the lessons from this study can be extended to higher-point energy correlators projected to the largest angular separation.

As shown in detail in \cite{Andres:2023xwr}, the EEC in Eq.~\eqref{eq:distribution}  can be written for a heavy-ion quark-jet, produced at a fixed hard scale $Q$, which traverses a QCD medium along a trajectory paramtrized by $\xi$, as
\begin{align}
    &\frac{\td \Sigma_{\xi}(Q)}{\td R_L} = \frac{1}{\sigma}\int \td z  \left(g^{(1)}(R_L,\As) + F^{\rm med}_{\xi}(z,R_L)  \right) \frac{\td \hat\sigma^{\rm vac}_{qg}}{\td R_L \td z }  z (1-z) \left(1  + \mathcal{O}\left(\frac{\bar \mu_{\rm s} }{Q} \right) + \mathcal{O}\left(\frac{\Lambda_{\mathrm{QCD}}}{R_L ~ Q}\right) \right) \,,  
    \label{eq:eec}
\end{align}
in which, for an initiating quark with energy $E$, $F_{\xi}^{\rm med}$ is the medium modification along the quark's trajectory to the vacuum partonic cross-section for the $q \to qg$ splitting $\td\hat\sigma^{\rm vac}_{qg}$, with $z$ being the energy fraction $z = E_g /E$. $\sigma(Q)$ denotes the integrated cross-section. The vacuum resummation is contained in $g^{(1)}$, i.e. $g^{(1)} = R_L^{\gamma(3)} + \mathcal{O}(R_L)$ at fixed coupling.  We refer the interested reader to the appendix~A of ref.~\cite{Andres:2023xwr} for the full expression of $g^{(1)}$ with running coupling. We note that given that ~\eqref{eq:eec} was obtained for wide-cone jets produced at a fixed initial energy $E$ at central rapidity,  one can set $Q=E$. A completely equivalent expression to eq.~\eqref{eq:eec} is found for a heavy-ion gluon-jet. Hence, for a mixed sample, the quark and gluon expressions should be summed, each weighted by their corresponding $q/g$ fractions inherited from the initial state and hard process. 

Since \eqref{eq:eec} describes a perturbative wide-cone jet produced at a fixed scale, it can only be immediately applied to quark-jets tagged with a $\gamma/Z$~boson. In this work we extend the computation of the EEC spectrum to a sample of inclusive quark-tagged jets with a moderate jet radius. Since the $q/g$ fraction of the sample does not significantly change the qualitative features of the EEC spectrum, we use the simplifying assumption that the quark tagging is completely efficient. Just as in Eq.~\eqref{eq:eec}, it is theoretically simple but computationally more expensive to include a generalized $q/g$ fraction in the sample. The generalization towards inclusive (quark) jets is as follows: 
\begin{align}
    \frac{\td \Sigma (\tilde{Q})}{\td R_L} = \int \td Q \int \td \epsilon ~~ \delta(Q - \epsilon - \tilde{Q}) ~ \< P_{\xi}(\epsilon) ~ \frac{\td \Sigma^{\rm NP}_{\xi} (x Q)}{\td R_L}  \>_{\xi} \otimes \mathcal{H}_{q}(x, Q)\,,
    \label{eq:eecinclusivjet}
\end{align}
which we proceed to explain in the following. 

Firstly, in \eqref{eq:eecinclusivjet} we have introduced $\tilde{Q}$ as the quenched hard scale, representing the jet's energy after experiencing energy loss. This energy loss is encoded in the probability $P_{\xi}(\epsilon)$ for the jet to lose an amount of energy $\epsilon$ along the trajectory $\xi$. The calculation of $P_{\xi}(\epsilon)$ will be explained in detail section~\ref{sec:weights}. Next, $\td \Sigma^{\rm NP}_{\xi} (x Q)/\td R_L$ denotes the EEC of a heavy-ion quark-jet produced at a hard scale $xQ$, as defined in Eq.~\eqref{eq:eec}, but in which the perturbative computation of $\td \hat{\sigma}_{qg}^{\rm vac}$ and $g^{(1)}$ has been supplemented by a non-perturbative (NP) approach to model the transition to hadronization to be discussed in detail in section~\ref{sec:transition}. Finally, $\mathcal{H}_{q}(x,Q)$ represents the self-normalised hard function for the inclusive production of a quark jet, and $\otimes$ represents the convolution over the momentum fraction $x$ between the hard function and the scale of eq.~\eqref{eq:eec} as defined in \cite{Dixon:2019uzg}.\footnote{In the notation of \cite{Dixon:2019uzg}, $\td \Sigma^{\rm NP}_{\xi} (x Q)/\td R_L$ can be understood as the differential EEC jet function for a jet produced at the hard scale $\mu=xQ$ propagating through a QCD medium.} This hard function for jets at central rapidity can be expressed at leading order as
\begin{equation}
    \mathcal{H}_{q}(x, Q) = \delta(1-x) \int \td E_{q}~ \delta(Q-E_{q}) ~\frac{1}{\sigma_{q+X}} \frac{\td \sigma_{q+X}}{\td E_{q}}\, ,
\end{equation}
where $E_{q} = p_{T,q}$ is the partonic energy/$p_{T}$ of the quark produced at the hard scattering.

The bracket $\<\>_{\xi}$ in eq.~\eqref{eq:eecinclusivjet} represents the average over all possible emitter's trajectories within a given centrality class of a given hydrodynamic simulation describing the QGP evolution. In this average, each trajectory must be weighted by the probability of producing the emitter at its initial point $(x_0, y_0)$, which can be obtained from the nuclear overlap distribution of the incoming nuclei, as follows\footnote{We note that the production time of the emitter will be assumed to coincide with the initial proper time of the hydrodynamic simulation, as customary in the field. Exceptions to this assumption, particularly in the context of energy loss, can be found in \cite{Armesto:2009zi,Andres:2016iys,Andres:2019eus,Zigic:2019sth,Stojku:2020wkh,Li:2020kax,Andres:2022bql}.}
\beq
 w(x_0,y_0) = T_A(x_0,y_0)\,T_A(\v b-(x_0,y_0))\,,
 \label{eq:weight}
\eeq
where $T_A$ is 3-parameter Fermi distribution at a given impact parameter $\v b$. 

Hence, the average of a given function $A_{\xi}$ within a specific centrality class characterized by the impact parameter $\v b$, is given by
\beq
\langle A_{\xi} \rangle_{\xi} = \frac{1}{N}\int {\rm d}\phi\, {\rm d} x_0\,{\rm d}y_0 \, w(x_0,y_0)\, A_{\xi}\,,
\label{eq:average}
\eeq
where $N= 2\pi \int  {\rm d} x_0\,{\rm d}y_0 \, w(x_0,y_0)$. We note that in  $\td \Sigma^{\rm NP}_{\xi} (x Q)/\td R_L$ only $F_{\xi}^{\rm med}(z, R_L)$ depends on the emitter's trajectory $\xi$. Thus, in practice, $F_{\xi}^{\rm med} (z, R_L)$ and $P_{\xi}(\epsilon)$ are the only terms in \eqref{eq:eecinclusivjet} that need to be averaged.

In this manuscript, we will employ the smooth-averaged 2+1 viscous hydrodynamic model developed in~\cite{Luzum:2008cw,Luzum:2009sb}. This simulation employs an initial condition derived in the Kharzeev-Levin-Nardi (KLN) $k_T$-factorization Color-Glass-Condensate approach~\cite{Drescher:2006pi} and a shear viscosity to entropy density ratio fixed to $\eta/s = 0.16$. The simulation begins at an initial proper time of $\tau_{\rm hydro}=1\,\rm{fm/c}$ and employs an equation of state inspired by Lattice QCD calculations. The system is assumed to be in chemical equilibrium until it reaches a freeze-out temperature of $T_{\rm f} = 140$ MeV. Results will be presented for the 0-10$\%$ and  20-30$\%$ centrality class in $\sqrt{s_{\rm NN}} =5.02$ TeV Pb-Pb collisions at the LHC.  We note that we do not analyze the effect of event-by-even fluctuations, since given the definition of this observable \eqref{eq:distribution}, their impact is expected to be very mild. It would  very interesting to formulate a more differential version of eq.~\eqref{eq:distribution} such that there is explicit sensitivity to the medium geometry and/or fluctuations, but we consider this beyond the scope of this study.

Finally, we would like to emphasize that just as for eq.~\eqref{eq:eec}, an analogous expression to \eqref{eq:eecinclusivjet} can be found for gluon jets. Qualitatively, the EEC of a gluon jet is very similar to that of a quark jet, with the only significant difference being $\mathcal{O}(1)$ changes in the the size of the features in their EEC spectra due to their different color structure. In this work, we do not model the initial state and thus do not account for the quark/gluon fractions, focusing solely on an inclusive quark jet sample. Nevertheless, due to the equivalence of the observable between quark and gluon jet samples, our results are entirely sufficient for capturing the qualitative picture of heavy-ion EECs in inclusive (mixed) samples we aim to present. For more quantitative statements, it would be necessary to include the gluon channel and account for the quark/gluon fractions.

We now proceed to provide a detailed explanation of how the various components in eq.~\eqref{eq:eecinclusivjet} are obtained.

\subsection{Determination of $F_{\xi}^{\rm med}$}

The calculation of $F^{\rm med}_\xi$ keeping both the angle $R_L$ and the energy fraction $z$, as required by eq.~\eqref{eq:eec}, is far from straightforward, even when assuming the medium to be static. Currently, for the $q \rightarrow qg$ channel only partial results are available \cite{Apolinario:2014csa,Blaizot:2012fh,Sievert:2018imd}. A novel promising numerical method has been recently obtained \cite{Isaksen:2023nlr}, but it is presently limited to the $\gamma \rightarrow q \bar q$ channel within static media with a length of up to $L=2$ fm. Given our need to keep track of both the  energy fraction $z$ and angle $R_L$ of $q \rightarrow qg$ splittings within a realistic expanding QGP, we will consider two different approximations to be described in the following.

\subsubsection{Multiple Scattering Approach in Evolving Media}
\label{subsec:HO}

The first approximation consists of accounting for multiple in-medium scatterings within the BDMPS-Z pertubative QCD framework \cite{Baier:1996kr,Baier:1996sk,Zakharov:1996fv,Zakharov:1997uu}. Since we need to keep track of both $z$ and $R_L$, we cannot employ usual approaches  where, either the soft limit, in which $z\to 0$, is taken \cite{Wiedemann:2000za,Mehtar-Tani:2006vpj,Andres:2020vxs}, or $z$ is kept finite but the angular dependence is lost \cite{Zakharov:1996fv,Zakharov:1997uu,Jeon:2003gi,Caron-Huot:2010qjx,Schlichting:2021idr}. In our previous works on energy correlators \cite{Andres:2022ovj,Andres:2023xwr,Andres:2023ymw}, we used, instead, the semi-hard approximation developed in \cite{Dominguez:2019ges, Isaksen:2020npj}. In this setup, all partons are energetic enough to propagate following straight-line classical trajectories in coordinate space, allowing us to keep track of both the angle $R_L$ and the energy sharing $z$ of the splitting while resumming all multiple scatterings. However, this approach neglects any transverse momentum broadening, that is, any fluctuations in the transverse position of the daughter partons. In this manuscript, we relax this semi-hard approximation by accounting for the leading non-eikonal corrections due to broadening. The origin and steps to follow to compute such corrections were outlined, for the $\gamma \to q \bar q$ splitting, in the appendix~A of ref.~\cite{Dominguez:2019ges}. The complete derivation of these corrections for the medium-induced $q \to q g$ splitting will be presented in a forthcoming publication \cite{forwardcite}. For completeness, we provide below the expressions of  $F^{\rm med}_{\xi} (z, R_L)$ within this semi-hard approach including these broadening corrections
\begin{align}
    F^{\rm med}_{\xi} (z, R_L) = \frac {\pi}{z^2(1-z)^2E^2} &
    \int \frac{{\rm d}^2 \v k}{(2\pi)^2} \,\delta(|\v k|-z(1-z)ER_L)\,
    \text{Re} \int_0^L {\rm d}t  \, \Big[i \v k \cdot (\v k + i \,\partial_{\v u_1})  
    C^{(3)}\left( L, t; \v u_1 \right)  \nonumber\\
    +  & \int_t^{L} {\rm d } \bar t\, e^{-i \frac{\v k^2}{2z(1-z)E}(\bar t -t)} \v k^2 \, (\v k + i \,\partial_{\v u_1}) \cdot (\v k - i \,\partial_{\v {\bar u_2}}) \,C^{(4)}(L, \bar t, t; \v u_1, \v{\bar u_2}) \nonumber\\
    & \qqqquad\times\, C^{(3)}(\bar t, t; \v u_1)  \Big]_{\substack{\v {u}_1=0 \\\bar{\v {u}}_2=0}}\,,
    \label{eq:HOspectrum}
\end{align}
where  $E$ is the energy of the emitter, $\v k$ is the relative transverse momentum between the daughters, and $C^{(n)}$ are the $n$-particle correlators. For the specific case of $q \to qg $ splitting, $C^{(3)}$ and $C^{(4)}$ can be written in terms of dipoles $S$ and quadrupole $Q$ as follows
\beq
C^{(3)}(\bar t, t ; \v u_1)= S(\bar t, t;[\v u_{\rm cl}])\,S(\bar t, t; [(1-z)\v u_{\rm cl}])\,,
\eeq
\beq
C^{(4)} (L, \bar t, t ; \v u_1, \v{\bar u_2})= Q(L, \bar t; [\v u_{\rm cl}, \v{\bar u}_{\rm cl}])\, S(L, \bar t; [(1-z)(\v u_{\rm cl}-\v{\bar u}_{\rm cl})])\,,
\eeq
with $\v{u}_{\rm cl}(s) = \v{u}_1 + \frac{\v{k}}{z(1-z)E}(s-t)$ and $\bar{\v{u}}_{\rm cl}(s) = \bar{\v{u}}_2 + \frac{\v{k}}{z(1-z)E}(s-\bar t)$. 

The dipole $S$ is given by
\beq
    S(t, \bar t, [\v r])= \exp \left[ -\frac{1}{2} \int_t^{\bar t} {\rm d}s\, n(s) \,\sigma(\v r(s)) \right]\approx \exp \left[ -\frac{1}{4} \int_t^{\bar t} {\rm d}s\, \hat{q}(s) \,\v r^2(s) \right]\,,
\eeq
where we have approximated the product of the linear density of scattering centers $n$ and the dipole cross-section $\sigma$ by $n(t)\sigma(\v r(t)) \approx \hat{q}(t)\v r^2(t)/2$. This is the so-called harmonic oscillator approximation (HO), which accounts for multiple soft interactions.

The quadrupole $Q$ can be written as
\begin{align}
Q(L, \bar t; [\v u_{\rm cl}, \v{\bar u}_{\rm cl}])=\,& S(L, \bar t; [z(\v u_{\rm cl}-\v{\bar u}_{\rm cl})]) \,S(L, \bar t; [(1-z)(\v u_{\rm cl}-\v{\bar u}_{\rm cl})]) + \int_{\bar t}^{L} {\rm d}{s} \,S(L, s; [z(\v u_{\rm cl}-\v{\bar u}_{\rm cl})]) \nonumber \\ 
& \times \,S(L, s; [(1-z)(\v u_{\rm cl}-\v{\bar u}_{\rm cl})]) \, T(s)  \, S(s, \bar t;[\v u_{\rm cl}])\,S(s, \bar t; [\v{\bar u}_{\rm cl}])\,,
\end{align}
with the transition matrix $T$ in the HO approximation being
\beq
    T(s) = -\frac{\hat{q}(s)}{2} z(1-z) \left ( \v u_1 -  \v{\bar u_2} + \frac{\v{k}}{z(1-z)E}(\bar t -t) \right)^2 \,.
\eeq

It becomes now clear that the information about the medium entering \eqref{eq:HOspectrum} is encoded in the length $L$  of the trajectory $\xi(t)$, to be sampled from the hydrodynamical simulation, and the  local time-dependent  jet quenching coefficient $\hat{q}(t)$. To proceed, we need to establish the relationship between 
$\hat q (t)$ and the local value of the temperature along the trajectory $T(\xi)$, which we set to \cite{Baier:2002tc}
\beq
\hat{q}(t) = k_{\rm HO}\, T^3(\xi(t))\,.\label{eq:kho}
\eeq
Hence, $k_{\rm HO}$ remains as  the only free parameter in this formalism  to compute $F^{\rm med}_\xi$ --- and the two-point correlator in \eqref{eq:eec}.

\subsubsection{Single Scattering Approach (GLV) in Evolving Media}
\label{subsec:GLV}
We now evaluate the medium-induced two-particle cross-section in an opacity expansion, which defines a series  in terms of the number of scatterings between the probe and the QCD medium. At finite $z$, the $q \to qg$ splitting was initially obtained at first order in opacity in \cite{Ovanesyan:2011kn}, and a method to recursively derive higher orders was developed in \cite{Sievert:2018imd,Sievert:2019cwq}. However, due to the rapid increase in complexity with the number of scatterings, these expressions become impractical even for static media, and thus  we restrict our analysis to the single scattering results in \cite{Ovanesyan:2011kn}. The main advantage of this first order in opacity formalism is its avoidance of the semi-hard approximation used in the multiple scattering approach presented in the previous section. The main drawback is that it yields incorrect results in  the phase space regions where coherence effects among multiple scatterings are expected to be important. Here, we present the expressions at first order in opacity in \cite{Ovanesyan:2011kn} adapted to our notation and applicable to non-static media\footnote{We note that this is precisely the same single scattering approach employed in \cite{Andres:2023xwr}, but generalized here to dynamically evolving media.}
\begin{align}
F^{\rm med}_{\xi}(z,R_L) = &\,\frac{2\pi\,}{z(1-z)E} \int 
\frac{{\rm d}^2 \v k}{(2\pi)^2}\, \delta\left(|\v k|-z(1-z)ER_L\right)
\int_0^L {\rm d}t \,n(t)\int \frac{{\rm d}^2 \v q}{(2\pi)^2}\,  V_{\rm Y}(\v q,t) \nonumber\\
&\times\v k^2\left[\frac{\v k + z \v q}{(\v k + z \v q)^2}\cdot \left(\frac{2C_{\rm F}}{N_c}\frac{\v k + z \v q}{(\v k + z \v q)^2} - \frac{\v k -  (1-z) \v q}{(\v k -  (1-z) \v q)^2} + \frac{1}{N_c^2}\frac{\v k}{\v k^2}\right) \left[1-\cos{(\Omega_{1} t)}\right] 
\right.  \nonumber\\
&\quad +  \frac{\v k -  (1-z) \v q}{(\v k -  (1-z) \v q)^2} \cdot \left(\frac{2(\v k -  (1-z) \v q)}{(\v k -  (1-z) \v q)^2}-\frac{\v k + z \v q}{(\v k + z \v q)^2} - \frac{\v k}{\v k^2}\right) \left[1-\cos{(\Omega_{2} t)} \right] \nonumber \\
&\quad +\frac{(\v k + z \v q)\cdot(\v k -  (1-z) \v q)}{(\v k + z \v q)^2 (\v k -  (1-z) \v q)^2}\left[1-\cos{(\Omega_{3} t)} \right] \nonumber \\
&\left.\quad + \frac{\v k}{\v k^2}\cdot \left(\frac{\v k - \v q}{(\v k - \v q)^2}- \frac{\v k}{\v k^2}\right)\left[1-\cos{(\Omega_{4} t)} \right] - \frac{\v k\cdot(\v k - \v q)}{\v k^2(\v k - \v q)^2} \left[1-\cos{(\Omega_{5} t)} \right]
\right]\,, 
\label{eq:GLVspectrum}
\end{align}
where $\v k$ is the relative transverse momentum of the daughter partons and 
\begin{align}
& \Omega_{1} = \frac{(\v k + z \v q)^2}{2z(1-z)E}\,,\quad
\Omega_{2} = \frac{(\v k -  (1-z) \v q)^2}{2z(1-z)E} \,, \quad
\Omega_{3} = \frac{(\v k -  (1-z) \v q)^2 -(\v k + z \v q)^2}{2z(1-z)E} \,,\nonumber\\
& \Omega_{4} = \frac{\v k^2}{2z(1-z)E} \,,\quad
 \Omega_{5} = \frac{\v k^2- (\v k - \v q)^2}{2z(1-z)E} \,.
\end{align}
In \eqref{eq:GLVspectrum} the information about the medium is encoded in: the length $L$ of the trajectory sampled from the hydrodynamical description of the medium, appearing as the the upper limit of the  integral over $t$; the time dependent linear density $n(t)$; and  the Yukawa collision rate $V_{\rm Y}(\v q, t)$, implemented in the Gyulassy-Wang model \cite{Gyulassy:1993hr} through 
\beq
V_{\mathrm{Y}}(\v q,t) = \frac{8 \pi \mu^2(t)}{\left(\v q^2 + \mu^2(t)\right)^2}\,,
\label{eq:V_yuk}
\eeq
where $\mu^2(t)$ is the time-dependent screening mass. Therefore, to compute $F^{\rm med}_{\xi}$ along a given path sampled from the hydrodynamic simulation, one needs to establish the relationship of the linear density of scattering centers and screening mass with the local properties of the medium. As customary in the field, we set this relationship  along the emitter's trajectory  $\xi(t)$ to be:
\beq
n(t) = k_{\rm GLV} T(\xi(t))\,,  
\qquad  {\rm and } \qquad
\mu^2(t) = 6\pi\alpha_s\, T^2(\xi(t))\,, 
\label{eq:hydro_GLV}
\eeq
where $T(\xi(t))$ is the temperature along the emitter's path  read from a hydrodynamic model and $\alpha_s$ is the coupling strength of the in-medium scatterings, which we will set in the following to $\alpha_s = 0.3$. The proportionality of the screening mass to the temperature in \eqref{eq:hydro_GLV} follows the leading order result in the hard thermal loop perturbation theory \cite{Aurenche:2002pd}. Regarding the linear density $n(t)$, we recall that its use in this formalism  arises  from the assumption of homogeneity in the transverse plane, which leads to the replacement of the volume density $\rho$ by the linear density $n(t)=\rho(t)\sigma_{el}$, with $\sigma_{el}$ being the total elastic cross-section, which is proportional to $\mu^2$. Based on the Stefan-Boltzmann limit of the QCD equation of state, the volume density is proportional to $T^3$, and thus, the linear density is proportional to $T$, as in \eqref{eq:hydro_GLV}. Consequently, the proportionality of the linear density to the temperature, encoded in $k_{\rm GLV}$, is the only free parameter in the computation of  $F^{\rm med}_\xi$ within this first opacity formalism  --- and, thus in the determination of the two-point correlator in \eqref{eq:eec}.

%%%%%%%%%%%%%%%%%%%%%%%%%%%%%%%%%%%%%%%%%%%%%%%%%
\subsection{Hadronization Transition}
\label{sec:transition}
%%%%%%%%%%%%%%%%%%%%%%%%%%%%%%%%%%%%%%%%%%%%%%%%%

One of the most significant effects of selection bias on a sample of jets produced in heavy-ion collisions is the lack of control over the hard scale, $Q$, (i.e. the initial jet energy  $E$) that characterizes the substructure within the jet. In the energy correlator spectrum, features emerge at angles fixed by the ratio of the relevant dynamical scales against the hard scale. Given that this initial hard scale is unknown in heavy-ion inclusive samples due to energy loss, selection bias can re-scale the shape of the features present in the EEC spectrum causing the AA to pp ratio depart from the flat behaviour one would naively expect from universal dynamics. One of the most characteristic features in the EEC spectrum is the separation between non-perturbative physics and perturbative physics at an angle $\theta_{{\rm NP}}$. To account for the effect of selection bias over the complete EEC spectrum,  we are thus motivated to introduce a simple model which can qualitatively describe the transition to confinement at $\theta_{{\rm NP}}$.

The simplest description of the hadronization transition is found by smoothly interpolating the deeply non-perturbative and the perturbative regions of the vacuum EEC spectrum. In the deeply non-perturbative region, the EEC spectrum obeys an area law corresponding to the trivial correlations of a freely propagating hadron gas,
\begin{align}
    \frac{\td \Sigma}{\td R_L}\bigg|_{R_L \ll \theta_{{\rm NP}}} \sim R_L \,.
\end{align}
In contrast, the deeply perturbative part of the vacuum EEC spectrum obeys a scaling operator product expansion scaling law
\begin{align}
    \frac{\td \Sigma}{\td R_L}\bigg|_{\theta_{{\rm NP}} \ll R_L \ll 1} \sim R_L^{-1 + \mathcal{O}(\As)}\,.
\end{align}
An analytic function that interpolates between these two oppositely diverging limits can be obtained by expanding perturbatively using a Pad\'e approximant:
\begin{align}
    \frac{\td \Sigma}{\td R_L}\bigg|_{R_L< 1} = \frac{\sum_{i} A_i R_L^{i}}{\sum_{j} B_i R_L^{j}}\,.
\end{align}
The lowest order term in the expansion which describes the two limits is\footnote{We note that the linear non-perturbative power corrections to the EEC spectrum, which have been computed in the literature \cite{Schindler:2023cww}, appear in the expansion of the Pad\'{e} approximant as sums of the $B_3$ term. Therefore, any model introduced to describe just the $B_{0}$ and $B_{2}$ dependence will not recreate the linear power corrections.} 
\begin{align}
    \frac{\td \Sigma}{\td R_L}\bigg|_{R_L < 1} = \frac{R_L}{B_0 + B_2 R_L^2}\,, \qquad \text{with} \qquad \frac{B_0}{B_2} \sim \theta_{\rm NP}^2\,.
\end{align}
Our aim is thus to modify our matrix elements to reproduce this Pad\'{e} approximant for the  EEC spectrum without modifying its known operator product expansion scaling in the deeply perturbative region. This behavior can be achieved by considering the functional form of the cross-section for radiation off a massive particle with mass $m$ and initial energy $E$ in the small-angle limit \cite{Dokshitzer:1991fd}
\begin{align}
   \frac{\td \hat{\sigma}_{qg} }{ \td R_L}\bigg|_{R_L < 1} \sim \frac{R_L}{R_L^2 + \Theta^2}\,, \qquad \text{with} \qquad \Theta = \frac{m}{ E}\,.
\end{align}
We note that including small propagator masses in the computation of the EEC does not alter the leading terms in the resummation of the operator product scaling in the perturbative region where $R_{L}\gg m/Q$. 

These observations motivate us to introduce a simple model for the non-perturbative physics in the EEC spectrum by regulating propagator denominators by giving each particle a small off-shell mass. Additionally, the running coupling, which behaves as $\As(R_{L} Q)$ in the EEC spectrum, needs to be supplemented with the analytically continued running coupling, so that the Landau pole is regulated \cite{Prosperi:2006hx}. We note that this model is  functionally similar to that which was introduced to compute small non-perturbative power corrections to energy flow observables by Dokshitzer, Marchesini and Webber \cite{Dokshitzer:1995qm}. In the case of \cite{Dokshitzer:1995qm}, a small effective gluon mass enters via dispersion relations which also necessitate the usage of the analytically extended QCD coupling. However, in this work, we are applying a simplified version of this approach beyond the regime for which it was originally introduced,  aiming to capture non-perturbative dynamics that extends beyond power corrections to the spectrum.

We find that a off-shell mass regulator of $m= 1200 \pm 200$~MeV provides a reasonable qualitative description of p-p EEC presented in the literature \cite{Lee:2022ige,CMS:2024mlf}. However, we note that the model noticeably underestimates the amplitude of the hadronization peak around $\theta_{\rm NP}$. This discrepancy could be due to the absence of positive linear power corrections, $\mathcal{O}(\Lambda_{\rm QCD}/R_L Q)$, in our approach. Positive linear corrections, identified when the complete dispersive approach of Dokshitzer et al. is applied and confirmed by other analyses \cite{Belitsky:2001ij,Korchemsky:1999kt,Korchemsky:1997sy,Korchemsky:1994is,Schindler:2023cww,Lee:2024esz,Chen:2024nyc}, increase the amplitude of the EEC spectrum to the right of the peak region.  Modelling these linear terms goes beyond the scope of the current manuscript.

We apply this approach to describe the non-perturbative contribution to the EEC spectrum both  in p-p and Pb-Pb collisions. Our main assumption is that the fundamental dynamics underlying the hadronization features is universal. Specifically, we assume that in heavy-ion collisions, the medium does not generate an additional screening mass in the partons' propagators and the basic structure for the parton-to-hadron fragmentation is unaffected by the presence of the medium. However, within the medium, partons do lose energy before undergoing vacuum-like fragmentation into hadrons. As a consequence of this energy loss, reconstructed jets of given energy do not correspond to the same jet population in heavy-ion and p-p collisions. This is expected to result in the non-perturbative transition in the heavy-ion inclusive EEC appearing shifted toward lower $R_L$-values compared to the p-p case. 

\subsection{Energy Loss}
\label{sec:weights}

In this section, we focus on the impact of energy loss on energy correlators observables. We start from two observations. Firstly, the hadronization transition is a local very small-angle feature that does not depend on the high-scale processes which lead to broad structures such as the jet width. Secondly, it was previously argued in \cite{Andres:2023xwr} that the effects of energy-loss are formally sub-leading in the large-angle region of the (collinear) EEC. This was also numerically shown with a quenching model in \cite{Barata:2023bhh}. This means that the effect of the bias towards selecting narrower jets in heavy-ion collisions, which may have undergone less energy loss than broader jets, is expected to be subelading (and very mild) in these observables. Following these observations, a good understanding of the average effect of energy loss on the hadronization transition in the EEC spectrum can be achieved from understanding only the extrinsic properties of the jet sample --- in particular the distribution of the jet energies. For this purpose, we make use of the quenching weight (QW) formalism originally proposed in \cite{Baier:2001yt,Salgado:2003gb,Salgado:2002cd}.\footnote{We note that the quenching weights can be formally derived taking the soft limit of the rate equations for independent gluon emissions in \cite{Jeon:2003gi,Turbide:2005fk,Blaizot:2013vha}.}

The quenching weight $P_{\xi}(\epsilon)$ in \eqref{eq:eecinclusivjet}  encodes the probability that a highly energetic parton loses an amount of energy $\epsilon$ along its trajectory $\xi$ due to the radiation of an arbitrary number of independent medium-induced soft gluons. It can be written as
\beq
P_{\xi}(\epsilon)= 
\sum_{N=0}^\infty \frac{1}{N!} \,
\prod_{i=1}^N \,
\left [\int {\rm d}\omega_i \,\frac{{\rm d}I_{\xi}}{ {\rm d} \omega} \right]
\,\delta \left(\epsilon -\sum_{i=1}^{N} \omega_i \right)
\exp \left[-\int_0^\infty {\rm d}\omega \frac{{\rm d}I_{\xi}}{{\rm d}\omega} \right]\,,
\label{eq:QWs}
\eeq
where ${\rm d}I_{\xi}/{\rm d}\omega$ is the well-known medium-induced \textit{soft} gluon emission spectrum  ($z \to 0$ and $E \to \infty$ with $\omega= zE$ finite) within the HO approximation along the fixed trajectory $\xi$ \cite{Baier:1996kr,Baier:1996sk,Zakharov:1996fv,Zakharov:1997uu}. For the full expressions of the HO medium-induced spectrum in the soft limit see, for instance, the appendix~A of \cite{Salgado:2003gb}.

This quenching weight approach relies on two main assumptions: multiple soft gluon emissions occur independently, and jets lose energy as a single color source which then fragments as in the vacuum. These assumptions are strongly supported by theoretical calculations of quantum mechanical interference effects among multiple in-medium emissions. Within the color coherence picture of in-medium emissions, the fragmentation of high-energy partons remains unmodified if color coherence is maintained, which is expected to occur at small enough angles \cite{Mehtar-Tani:2010ebp,Mehtar-Tani:2011hma,Casalderrey-Solana:2011ule,Mehtar-Tani:2012mfa}. At the angular scales where non-perturbative effects become important, color coherence is thus expected to hold,  implying that the totally coherent scenario provides a reasonable approximation for energy loss within the hadronization transition of  the EEC. Furthermore, the independent gluon emission assumption is now supported by recent efforts  \cite{Arnold:2022mby, Arnold:2023qwi}  to compute the double gluonic splittings $g \to ggg$, which suggest that the quantum overlap among multiple splittings should be small. Corrections to this totally coherent QW approach to energy loss introduce a non-trivial dependence on the angle of the emissions and can in principle be computed  \cite{Mehtar-Tani:2017web,Mehtar-Tani:2017ypq}, but their impact on the energy correlator is expected to be mild.

The quenching weights \eqref{eq:QWs} depend on the local jet quenching coefficient $\hat{q}(t)$ that enters the calculation of the  BDMPS-Z soft spectrum ${\rm d}I_{\xi}/{\rm d}\omega$ along the trajectory $\xi$.  This jet transport coefficient can be expressed in terms of the local temperature as $\hat{q}(t)=K_{\rm soft}T^3(\xi(t))$.\footnote{
While the definitions of $K_{\rm soft}$ and $k_{\rm HO}$ in eq.~\eqref{eq:kho}  may appear similar, it is crucial to recognize that they represent distinct parameters. They are used under vastly different approximations (soft and semi-hard, respectively), and therefore absorb corrections of different types when determined through fitting to data.} Therefore, to fully determine $P_{\xi} (\epsilon)$, and thus estimate the effect of energy loss in the inclusive EEC \eqref{eq:eecinclusivjet}, we need first to extract the value of $K_{\rm soft}$.  In order to do so, we will perform a fit to the nuclear modification factor $R_{\rm AA}$ of charged hadrons with $p_T> 10$ GeV  in $\sqrt{s_{\rm NN}}=5.02$ TeV Pb-Pb collisions at the LHC \cite{ATLAS:2022kqu}.  We note that our main goal is not to achieve an accurate description of the experimental data, but rather to obtain an estimate $K_{\rm soft}$, which will allow us to  provide a qualitative picture of the effect of energy loss on inclusive heavy-ion EECs observables. Consequently, we will use a simplified partonic model for our calculation of the $R_{\rm AA}$.

%%%%%%%%%%%%%%%%%%%%%%%%%%%%%%%%%%%%%%%%%%%%%%%%%

We start by approximating the inclusive cross-section to produce a high energy quark as
\beq
{\rm d}\sigma_{q+X}/{\rm d}p_{T,q} \sim 1/p_{T,q}^n\,,
\eeq
so the partonic inclusive suppression along a fixed trajectory $\xi$ can be written as
\beq
\left. \frac{{\rm d}\sigma^{{\rm AA} }(p_T)/{\rm d}p_T}
{{\rm d}\sigma^{{\rm pp}}(p_T)/{\rm d}p_T} \right|_{\xi}= \int {\rm d}\epsilon \, P_{\xi} (\epsilon)
\left (\frac{p_T + \epsilon}{p_T} \right)^n\,,
\label{eq:quenching_factor}
\eeq
where we have defined $p_T\equiv p_{T,q}$.

Our estimate for the nuclear modification factor $R_{\rm AA}$ in a given centrality class will be obtained by averaging \eqref{eq:quenching_factor}
according to \eqref{eq:average} over the emitter' trajectories within that centrality class. We consider several values of our fitting parameter $K_{ \rm soft}$ and perform a $\chi^2$-fit to the $R_{\rm AA}$ data in that centrality to determine its best value.
The results for the two centrality classes analyzed in this paper, 0-10$\%$ and  20-30$\%$, are  $K_{ \rm soft}=16$ and $K_{ \rm soft}=19$, respectively. Hence, in the results presented in the next section for  the inclusive EEC \eqref{eq:quenching_factor} for the 0-10$\%$ (20-30$\%$) centrality class in $\sqrt{s_{\rm NN}}=5.02$ TeV Pb-Pb collisions, $P_{\xi}(\epsilon)$ will be computed using $K_{ \rm soft}=16$ ($K_{ \rm soft}=19$).

%%%%%%%%%%%%%%%%%%%%%%%%%%%%%%%%%%%%%%%%%%%%%%%%%
%%%%%%%%%%%%%%%%%%%%%%%%%%%%%%%%%%%%%%%%%%%%%%%%%
\section{Results and Discussion}
\label{sec:results}
%%%%%%%%%%%%%%%%%%%%%%%%%%%%%%%%%%%%%%%%%%%%%%%%%

\begin{figure}[t]
\centering
\includegraphics[scale=0.33]{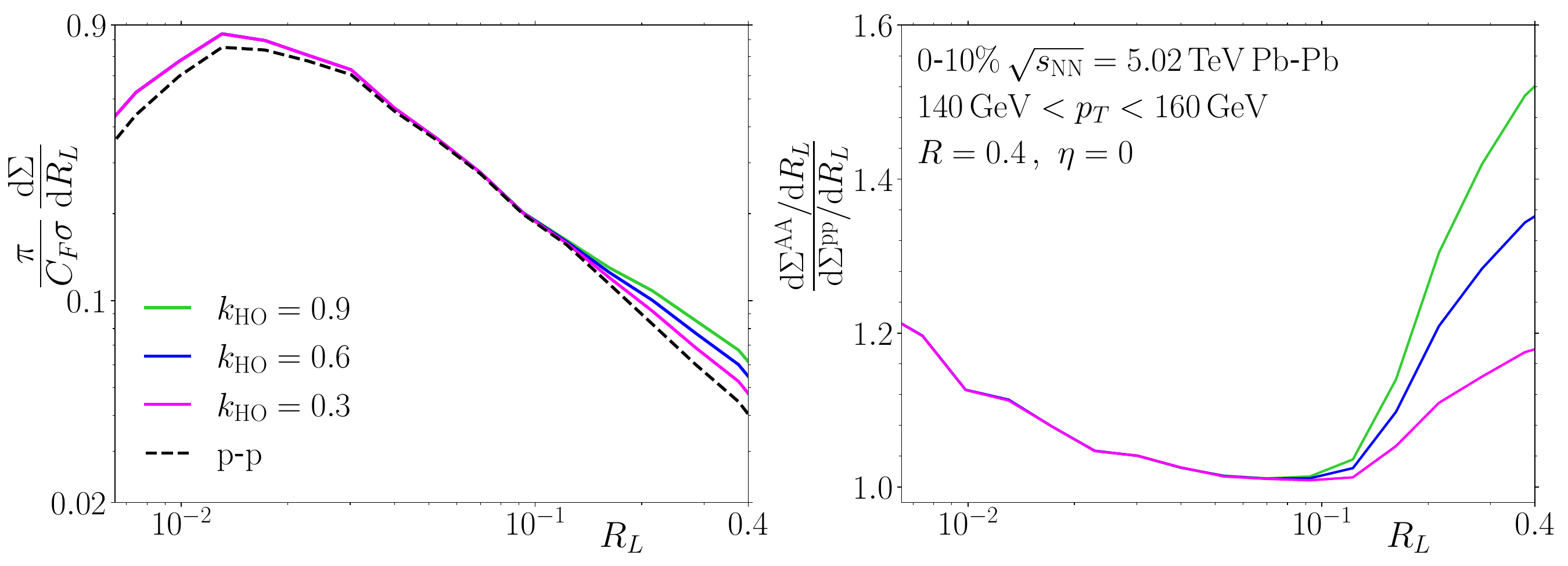}
\caption{Left panel: EEC for inclusive $140 <p_T <160$\,GeV  jets within the 0-10$\%$ centrality class in $\sqrt{s_{\rm NN}} =5.02$ TeV Pb-Pb collisions computed through eq.~\eqref{eq:eecinclusivjet} using the multiple scattering approach described in Section~\ref{subsec:HO} for several values of the free parameter $k_{\rm HO}$ (solid curves) compared to the p-p NLL (single-log accuracy) EEC (dashed black curve). Right panel: Ratio of the Pb-Pb EEC w.r.t the p-p EEC.}
\label{fig:HO3.2}
\end{figure}

In this section, we present numerical evaluations of the two-point energy correlator, as defined in eq.~\eqref{eq:eecinclusivjet}, focusing on jets with $140 < p_T < 160$ GeV at central rapidities in $\sqrt{s_{\rm NN}} = 5.02$ TeV Pb-Pb collisions at the LHC.\footnote{We note that throughout this section, $p_T$ represents the jet's energy, denoted in section~\ref{sec:inclusive} as $\tilde Q$.} These evaluations are conducted using the methods outlined in section~\ref{sec:inclusive}.\footnote{ Each figure is produced with the same numerical seed and therefore has correlated statistical fluctuations. }

The left panel of figure~\ref{fig:HO3.2} shows the EEC in inclusive $140 < p_T < 160$ GeV jet samples in p-p collisions (dashed line) and within the 0-10$\%$ centrality class in $\sqrt{s_{\rm NN}} =5.02$ TeV Pb-Pb collisions at the LHC (solid lines). The Pb-Pb EEC is computed through \eqref{eq:eecinclusivjet} with the medium modification $F^{\rm med}$ evaluated using the multiple scattering framework outlined in section~\ref{subsec:HO}, for several reasonable values of the free parameter $k_{\rm HO}$.  This free parameter can be fitted to any single-logarithmic collinear measurement (see section~\ref{sec:weights}), for instance a measurement of the two-point correlator in a particular $p_{T}$ bin and centrality, or a potential measurement of the one-point energy correlator. As reported in previous works \cite{Andres:2022ovj,Andres:2023ymw,Andres:2023xwr,Barata:2023bhh}, an enhancement due to medium-induced splittings can be observed at wide angles within the jet ($0.1 <R_L <0.4$). We highlight that the onset angle of this enhancement does not depend on the value of the free parameter $k_{\rm HO}$, as expected from the expressions presented in section~\ref{subsec:HO}. Compared to  \cite{Andres:2022ovj, Andres:2023xwr}, which reported results on the EEC of gamma-tagged jets using a multiple scattering formalism within a static brick, the medium-induced enhancement here is more modest, varying between 10$\%$-60$\%$ depending on the value of the free parameter $k_{\rm HO}$. This difference is due to the implementation of the hydrodynamical evolution of the medium and the generalization of the multiple scattering approach in  \cite{Andres:2022ovj, Andres:2023xwr}  to account for leading corrections due to transverse momentum broadening, as described in section~\ref{subsec:HO}.

At the small-angle end of the spectrum on the left panel of figure~\ref{fig:HO3.2}, we observe that the energy loss has shifted the hadronization transition of Pb-Pb jets towards smaller angles compared to that of p-p jets. This results in an increased ratio of the Pb-Pb to p-p EEC at small angles (see right panel of figure~\ref{fig:HO3.2}). This  effect has not yet been addressed in the literature, where analytic calculations have not studied the non-perturbative transition in Pb-Pb jets, and currently available Monte Carlo analyses \cite{Yang:2023dwc}, restricted to photon-tagged jets, show a suppression of the EEC in Pb-Pb jets relative to p-p in this region.\footnote{We note that \cite{Barata:2023bhh} includes a Monte Carlo study with JetMed \cite{Caucal:2018dla}. However, JetMed does not account for hadronization, and thus the observed  turnover in their figures is due to the shower cutoff rather than a model for confinement dynamics.} It would be very interesting to see if the small angle rise is observed in data, as it could be a strong discriminator between hadronization models in Pb-Pb collisions. 

It is important to note that the Pb-Pb  and p-p EEC spectra discussed throughout this section have been normalized to theoretical quantities which will not be experimentally accessible, particularly, the cross-section $\sigma$ to produce the quark jet from a partonic-state.  To make comparisons of the overall normalization against a specific experimental measurements, it would be necessary to include the complete hard function accounting for the initial state through (nuclear) parton distribution functions and the nuclear overlap function. Additionally, experimental efficiencies and cuts would need to be accounted for. Therefore, the overall normalization of these results should be considered a free parameter if a direct comparison against data is to be made.

\begin{figure}[t]
\centering
\includegraphics[scale=0.33]{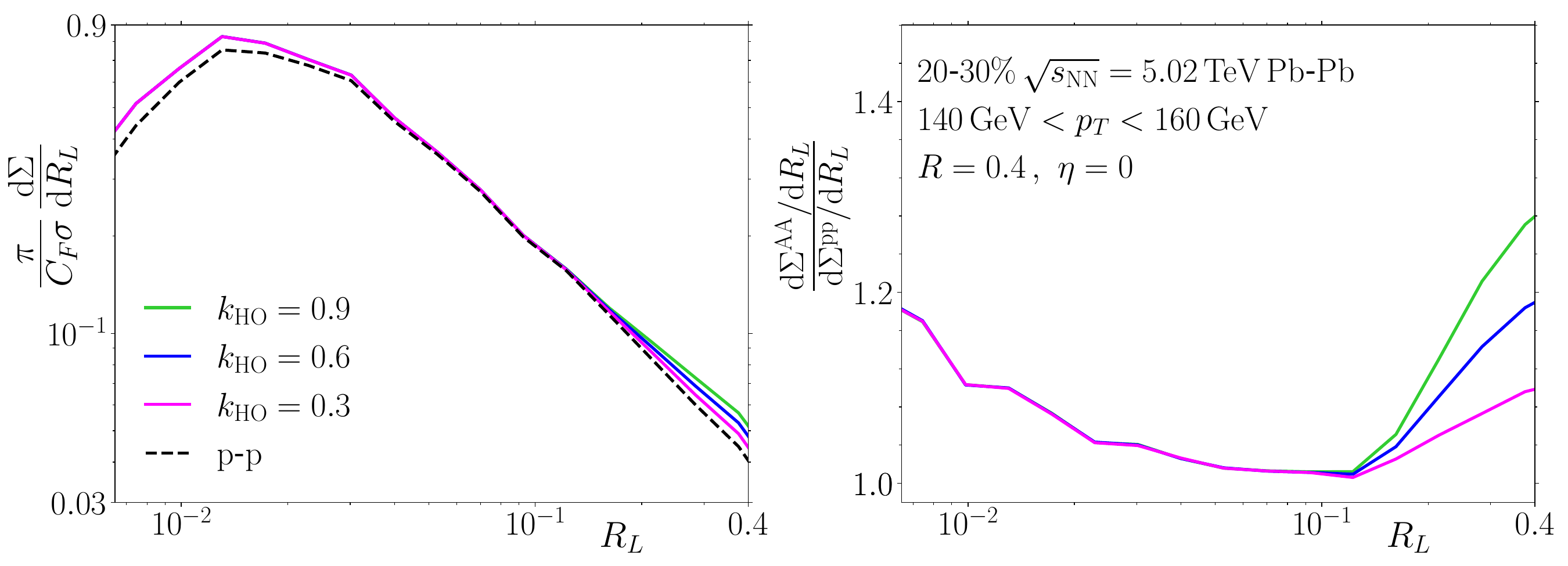}
\caption{Left panel: EEC for inclusive $140 <p_T <160$\,GeV  jets within the 20-30$\%$ centrality class in $\sqrt{s_{\rm NN}} =5.02$ TeV Pb-Pb collisions computed through eq.~\eqref{eq:eecinclusivjet} using the multiple scattering approach outlined in section~\ref{subsec:HO} for several values of the free parameter $k_{\rm HO}$ (solid curves) compared to the p-p NLL EEC (dashed black curve). Right panel: Ratio of the Pb-Pb EEC w.r.t the p-p EEC.}
\label{fig:HO7.2}
\end{figure}

Figure~\ref{fig:HO7.2} is analogous to figure~\ref{fig:HO3.2} but for the  20-30$\%$ centrality class in $\sqrt{s_{\rm NN}} =5.02$~TeV Pb-Pb collisions. As expected, the same basic features are observed, but their effects are reduced due to the more peripheral nature of the collisions.

\begin{figure}
\centering
\includegraphics[scale=0.33]{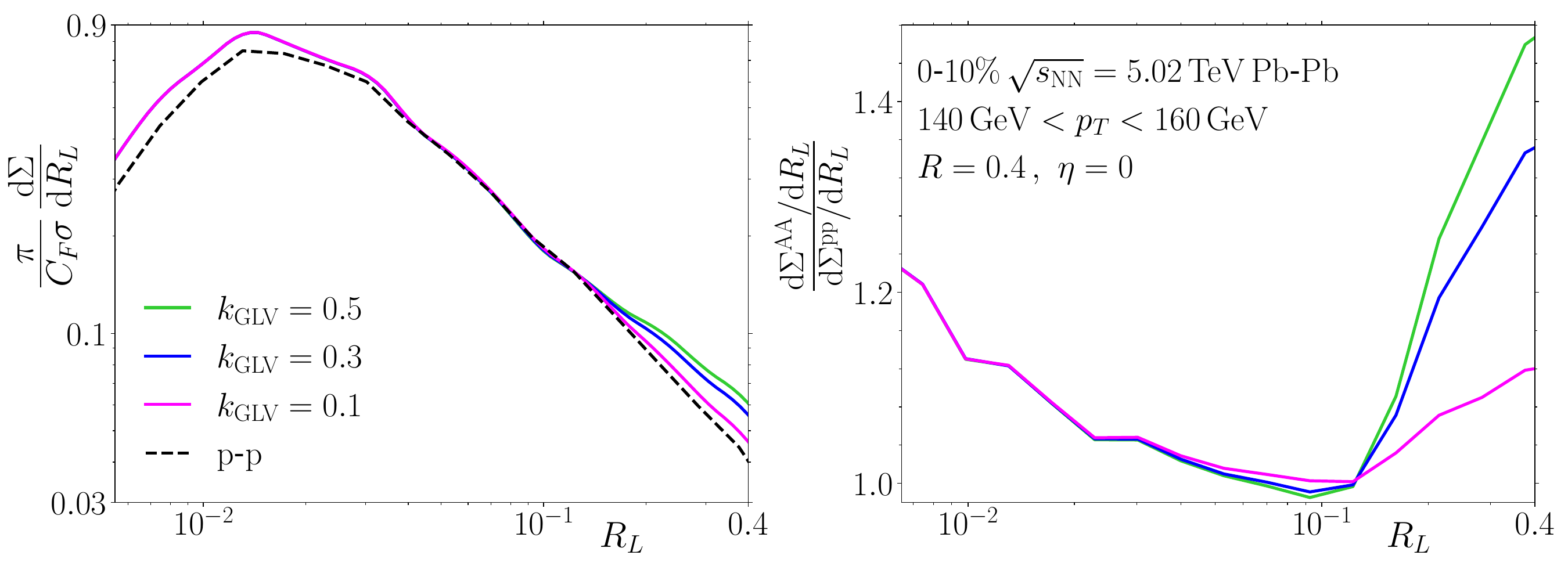}
\caption{Left panel: EEC for inclusive $140 <p_T <160$\,GeV  jets within the 0-10$\%$ centrality class in $\sqrt{s_{\rm NN}} =5.02$ TeV Pb-Pb collisions computed through eq.~\eqref{eq:eecinclusivjet} using the single scattering approach described in  \ref{subsec:GLV} for several values of the free parameter $k_{\rm GLV}$ (solid curves) compared to the p-p NLL EEC (dashed black curve). Right panel: Ratio of the Pb-Pb EEC w.r.t the p-p EEC.  }
\label{fig:GLV3.2}
\end{figure}

\begin{figure}
\centering
\includegraphics[scale=0.33]{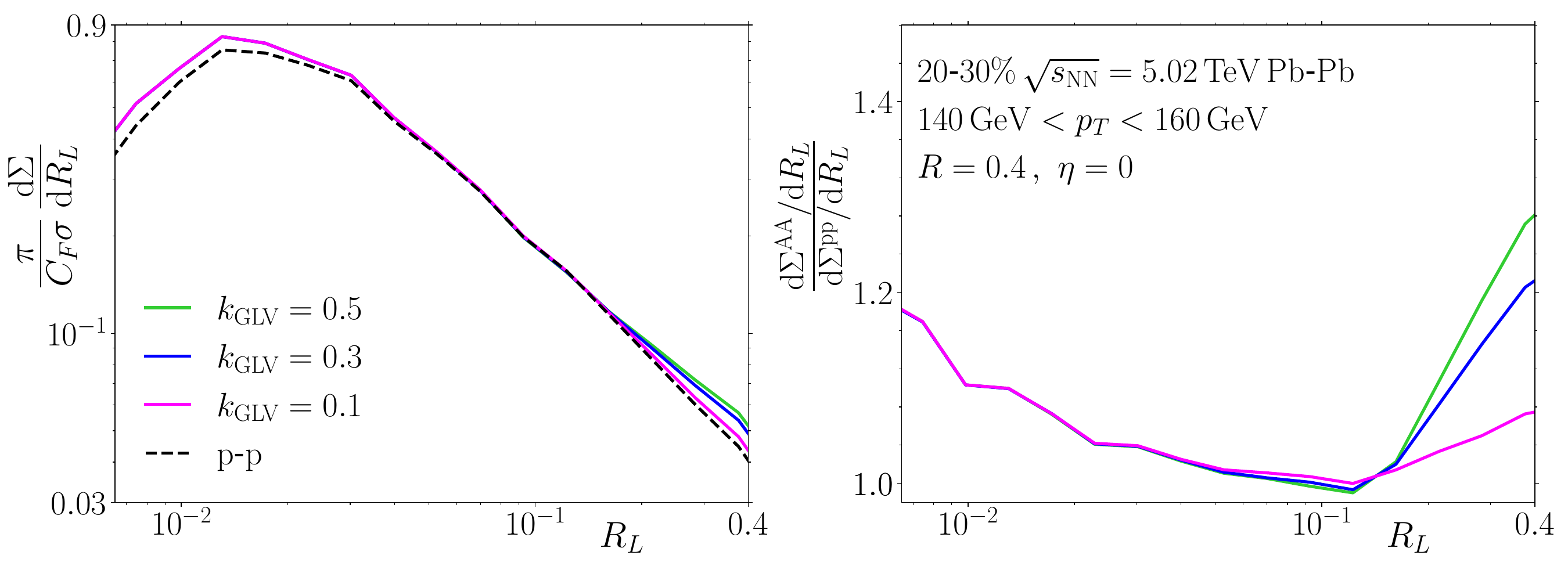}
\caption{Left panel: EEC for inclusive $140 <p_T <160$\,GeV  jets within the 20-30$\%$ centrality class in $\sqrt{s_{\rm NN}} =5.02$ TeV Pb-Pb collisions computed through eq.~\eqref{eq:eecinclusivjet} using the single scattering approach described in  \ref{subsec:GLV} for several values of the free parameter $k_{\rm GLV}$ (solid curves) compared to the p-p NLL EEC (dashed black curve). Right panel: Ratio of the Pb-Pb EEC w.r.t the p-p EEC.}
\label{fig:GLV7.2}
\end{figure}

We move now to figures \ref{fig:GLV3.2} and \ref{fig:GLV7.2} which display the results for the two-point energy correlator \eqref{eq:eecinclusivjet} with the medium-induced splittings computed within the single scattering formalism (GLV) described in section~\ref{subsec:GLV}.  Specifically, figure~\ref{fig:GLV3.2} shows the EEC  for inclusive jet samples in p-p (dashed) and 0-10$\%$ central $\sqrt{s_{\rm NN}} =5.02$ (solid), with $F^{\rm med}$ evaluated though \eqref{eq:GLVspectrum} for different values of the free parameter $k_{\rm GLV}$. The Pb-Pb EEC in figure~\ref{fig:GLV3.2}  shares the same fundamental characteristics as the Pb-Pb EEC evaluated with the multiple scattering approach shown in figure~\ref{fig:HO3.2}, albeit exhibiting a slightly different shape in the wide-angle enhancement.  It is important to note that while the single scattering approach used here is exactly the same as that employed for the EEC on gamma-tagged jets in \cite{Andres:2023xwr}, the enhancement at large angles in figure~\ref{fig:GLV3.2} is reduced compared to \cite{Andres:2023xwr}, due to the inclusion of a hydrodynamically expanding medium in this study.  

For completeness, we present in figure~\ref{fig:GLV7.2} the results on the EEC within the single scattering approach for the 20-30$\%$ centrality class  $\sqrt{s_{\rm NN}} =5.02$~TeV Pb-Pb collisions. As expected, the same fundamental features as in figure~\ref{fig:GLV3.2} are observed, but their effects are reduced due to the more peripheral nature of the collisions. 

Interestingly, the approaches to compute the medium-induced splittings presented in section~\ref{subsec:HO} and section~\ref{subsec:GLV} involve very different approximations.  The commonality of the main features in the EEC results from both analyses indicates a high degree of robustness in the understanding they provide. Nevertheless, it is still desirable to consider more complete calculations where multiple scatterings can be incorporated without the necessity of eikonal assumptions. New developments in that regard  will be consider in forthcoming publications \cite{forwardcite}.

\begin{figure}[t]
\centering
\includegraphics[scale=0.33]{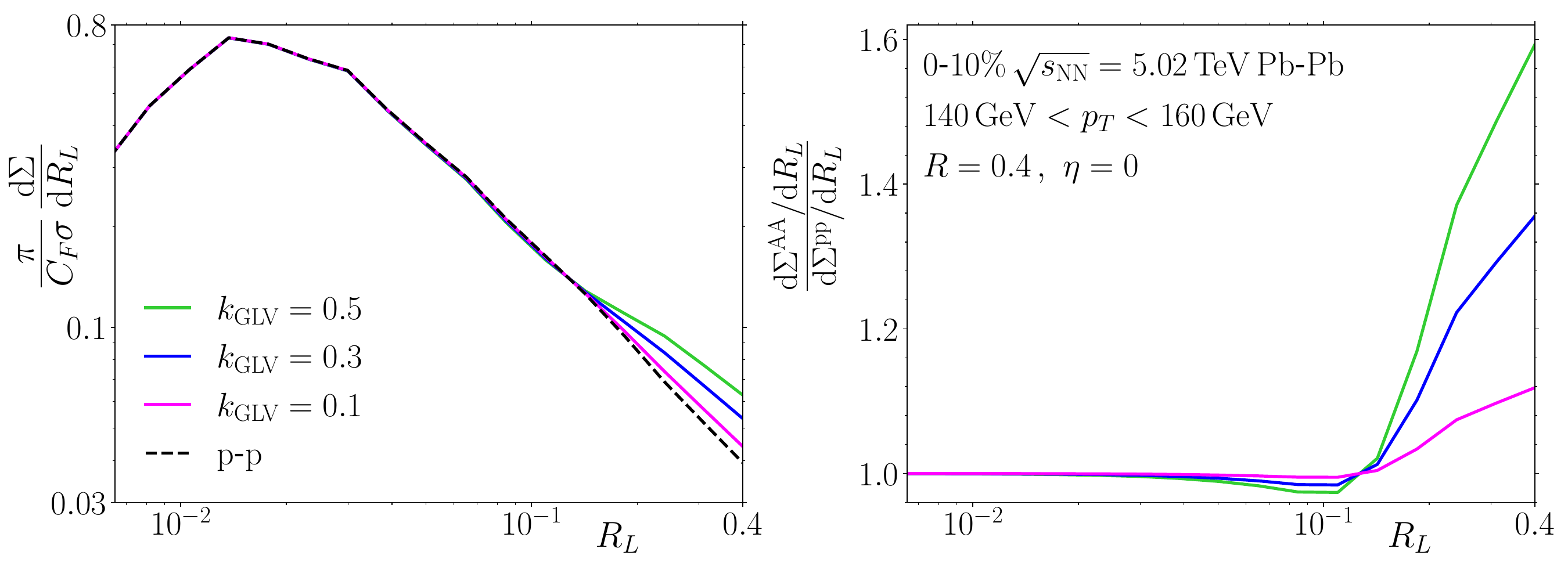}
\caption{Left panel: EEC for gamma-tagged $140 <p_T <160$\,GeV  jets within the 0-10$\%$ centrality class in $\sqrt{s_{\rm NN}} =5.02$ TeV Pb-Pb collisions computed through eq.~\eqref{eq:eec} using the single scattering approach described in  \ref{subsec:GLV} for several values of the free parameter $k_{\rm GLV}$ (solid curves) compared to the p-p NLL EEC (dashed black curve). Right panel: Ratio of the Pb-Pb EEC w.r.t the p-p EEC.  }
\label{fig:GLV3.2GammaTag}
\end{figure}

The last figure presented in this section is figure~\ref{fig:GLV3.2GammaTag}, which shows the results of the EEC for gamma-tagged jets with $140 < p_T < 160$ GeV in p-p and 0-10$\%$ central $\sqrt{s_{\rm NN}} =5.02$ TeV Pb-Pb collisions 
The Pb-Pb EEC in this figure is computed through \eqref{eq:eecinclusivjet} with $P(\epsilon)=\delta (\epsilon)$. The single scattering approach in \ref{subsec:GLV} is employed for the evaluation of $F^{\rm med}$. This figure serves as a reference to illustrate the impact of selection bias caused by energy loss on the earlier figures. Specifically, the absence of energy loss results in the hadronization transition occurring at the same angles for the p-p and Pb-Pb gamma-jet EECs in figure~\ref{fig:GLV3.2GammaTag}, resulting in the ratio of the Pb-Pb to p-p EEC at small angles being equal to 1 (see right panel of figure~\ref{fig:GLV3.2GammaTag}). Furthermore, the wide-angle ($0.1 < R_L <0.4$) enhancement is larger in these curves than in those of figure~\ref{fig:GLV3.2}.  This occurs because energy loss included in figure~\ref{fig:GLV3.2} has the effect of reducing the enhancement.  This reduction is due to selection bias, since the amplitude of the wide-angle enhancement weakly decreases with the hard jet energy scale $Q$. Additionally, it is worth noticing that a more sophisticated treatment of energy loss accounting for decoherence effects \cite{Mehtar-Tani:2017web,Mehtar-Tani:2017ypq}, which will cause energy loss to be dependent on the angle of the splitting, is expected to lead to a slightly greater reduction of the enhancement in  figure~\ref{fig:GLV3.2} in the $0.1 < R_L <0.4$ regime \cite{Barata:2023bhh}.

We further present in appendix~\ref{app:app} additional figures showing the two-point correlator in inclusive jet samples at central rapidities for different jet $p_{T}$ bins.

Finally, we note that we do not examine in this manuscript the effects of the medium back-reaction (or medium response) on the energy correlators. Although the back-reaction primarily involves soft hadrons and is therefore suppressed by the inclusiveness and energy weighting of the observable, it might still have a significant impact, as shown in \cite{Yang:2023dwc}. This impact is expected to occur in an overlapping angular region with the wide-angle enhancement from medium-induced radiation. In sum, these competing effects could obscure the interpretation of the medium's properties extracted from a measurement of this enhancement, motivating further theoretical work to better disentangle such sub-leading phenomena.

%%%%%%%%%%%%%%%%%%%%%%%%%%%%%%%%%%%%%%%%%%%%%%%%%

%%%%%%%%%%%%%%%%%%%%%%%%%%%%%%%%%%%%%%%%%%%%%%%%%
\section{Conclusions}
\label{sec:conc}
%%%%%%%%%%%%%%%%%%%%%%%%%%%%%%%%%%%%%%%%%%%%%%%%%
Energy correlator observables have recently emerged as a novel method for studying jet substructure in both proton-proton and heavy-ion collisions. Following successful measurements of energy correlators in proton-proton collisions at RHIC and at the LHC \cite{talk2, talk3, Tamis:2023guc, talk4, talk6, CMS:2024mlf}, alongside a robust theoretical understanding of their perturbative \cite{Dixon:2019uzg,Chen:2020vvp,Lee:2022ige,Chen:2023zlx} and non-perturbative \cite{Belitsky:2001ij,Korchemsky:1999kt,Korchemsky:1997sy,Korchemsky:1994is,Schindler:2023cww,Lee:2024esz,Chen:2024nyc} behavior, {and proposed phenomenological applications in p-p collisions \cite{Holguin:2022epo,Komiske:2022enw,Craft:2022kdo,Lee:2023npz,Holguin:2023bjf,Chen:2024nfl,Holguin:2024tkz}, the focus now shifts to their measurement in heavy-ion collisions. Given the significantly more complex environment of heavy-ion collisions, it is crucial to develop a physical understanding of how aspects of this environment may modify these observables.

In this manuscript, we have extended our work on boson-tagged jets \cite{Andres:2022ovj,Andres:2023xwr,Andres:2023ymw} to provide further analytical insight into the behavior of the two-point energy correlator in inclusive jet samples in heavy-ion collisions. Our goal is to offer a comprehensive qualitative description of the key physical effects present at leading order in heavy-ion jets within a controlled framework. This approach aims to facilitate interpretation of the first measurements of energy correlators in heavy-ion collisions, laying the groundwork for more precise quantitative calculations in the near future.

As compared to our previous calculations \cite{Andres:2022ovj,Andres:2023xwr,Andres:2023ymw}, we have incorporated: a realistic description of the expanding medium through a hydrodynamic simulation, broadening corrections to the semi-hard calculations of a quark splitting into a quark and a gluon undergoing multiple scatterings, selection bias due to energy loss, and a description of the non-perturbative transition via a simple new analytical model. These provide a much more realistic description of the heavy-ion environment, and allow us to develop a physical intuition of how each of these effects influences the EEC spectra. 

A crucial insight from this work is the description of how the bias associated to selecting jets with a given $p_T$ is imprinted in the energy correlators of heavy-ion inclusive jet measurements. As shown in section~\ref{sec:results}, this selection bias
shifts the location of the hadronization transition in the correlator spectrum toward lower angles in Pb-Pb compared to p-p. This leads to a non-trivial shape for the ratio between the Pb-Pb and p-p EEC distributions at small angles. We predict that this results in a rise of this ratio in this regime, provided the hadronization process is minimally affected by the heavy-ion environment. It will be interesting to see if this is observed in experimental data. This also further motivates the study of the energy correlators in $\gamma/Z+$ jet events, which we originally advocated for, since selecting jets based on the unmodified $p_T^B$ of the recoiling boson and on the momentum imbalance $x_J=p_T/p_T^B$ can significantly reduce the selection bias due to energy loss \cite{CMS:2024zjn,Brewer:2021hmh}. Additionally, this motivates further study of the hadronization transition beyond the simple analytical model presented here, a feature which has not yet seen a precision treatment in the literature.

Our results provide a key ingredient in the interpretation of the first measurements of the energy correlators in heavy-ion experiments, performed on an inclusive jet sample. We believe that the measurement of the energy correlators in heavy-ion collisions represents a milestone in the study of the QGP. As a direct measurement of a fundamental correlation function of a finite temperature relativistic gauge theory, it provides a key step in connecting experimental heavy-ion physics with fundamental quantum field theory. We look forward to detailed theoretical studies of this first result, as well as future more detailed experimental measurements.

%%%%%%%%%%%%%%%%%%%%%%%%%%%%%%%%%%%%%%%%%%%%%%%%%
\acknowledgments
We thank Jo\~ao Barata, Hannah Bossi, Raghav Kunnawalkam Elayavalli, Wenqing Fan, Laura Havener, Barbara Jacak, Yen-Jie Lee, Ananya Rai, Krishna Rajagopal, Jussi Viinikainen,  Xin-Nian Wang for useful discussions. I.M. is supported by startup funds from Yale University. J.H. is supported by the Leverhulme Trust as an Early Career Fellow. 
This work is supported by the GLUODYNAMICS project funded by the ``P2IO LabEx (ANR-10-LABX-0038)'' in the framework ``Investissements d’Avenir'' (ANR-11-IDEX-0003-01) managed by the Agence Nationale de la Recherche (ANR), France; by the European Research Council project ERC2018-ADG-835105 YoctoLHC; by Maria de Maeztu excellence unit grant CEX2023-001318Mand project PID2020-119632GB-I00 funded by MICIU/AEI/10.13039/501100011033; by ERDF/EU. It has received funding from Xunta de Galicia (CIGUS Network of Research Centres); and from the European Union’s Horizon 2020 research and innovation programme under grant agreement No. 824093. This work was supported by OE Portugal, Funda\c{c}\~{a}o para a Ci\^{e}ncia e a Tecnologia (FCT), I.P., project 2024.06117.CERN. CA acknowledges the financial support by the FCT under contract \newline 2023.07883.CEECIND. The authors would like to express special thanks to the Mainz Institute for Theoretical Physics (MITP) of the Cluster of Excellence PRISMA$^+$ (Project ID 390831469), for its hospitality and support.

\appendix

\section{Additional Figures}
\label{app:app}

In this appendix, we present additional figures showing the two-point correlator \eqref{eq:eecinclusivjet} computed using the single scattering formalism (GLV) described in \ref{subsec:GLV} for different jet $p_T$ bins
in 0-10$\%$ central $\sqrt{s_{\rm NN}} =5.02$ TeV Pb-Pb collisions  compared to the p-p result. Specifically, figure~\ref{fig:GLV3.2_100} displays results for $100 < p_T < 120$ GeV jets and figure~\ref{fig:GLV3.2_180} for $180 < p_T < 200$ GeV jets. We note that as the jet $p_T$ increases the effect of energy loss decreases, thereby reducing the jet selection bias. This trend is clearly observed by comparing the right panels of both figures in the small angular regime.

\begin{figure}[ht]
\centering
\includegraphics[scale=0.33]{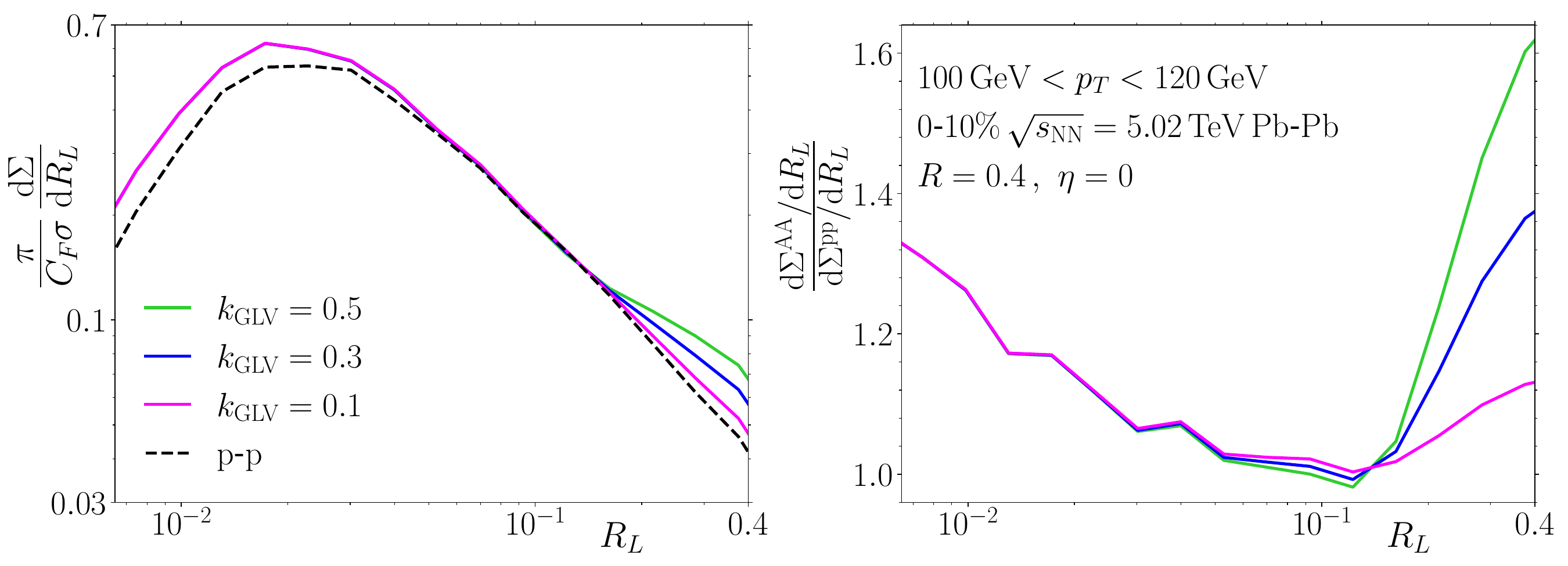}
\caption{Left panel: EEC for inclusive $100 <p_T <120$\,GeV  jets within the 0-10$\%$ centrality class in $\sqrt{s_{\rm NN}} =5.02$ TeV Pb-Pb collisions computed through eq.~\eqref{eq:eecinclusivjet} using the single scattering approach described in  \ref{subsec:GLV} for several values of the free parameter $k_{\rm GLV}$ (solid curves) compared to the p-p NLL EEC (dashed black curve). Right panel: Ratio of the Pb-Pb EEC w.r.t the p-p EEC.  }
\label{fig:GLV3.2_100}
\end{figure}

\begin{figure}[ht]
\centering
\includegraphics[scale=0.33]{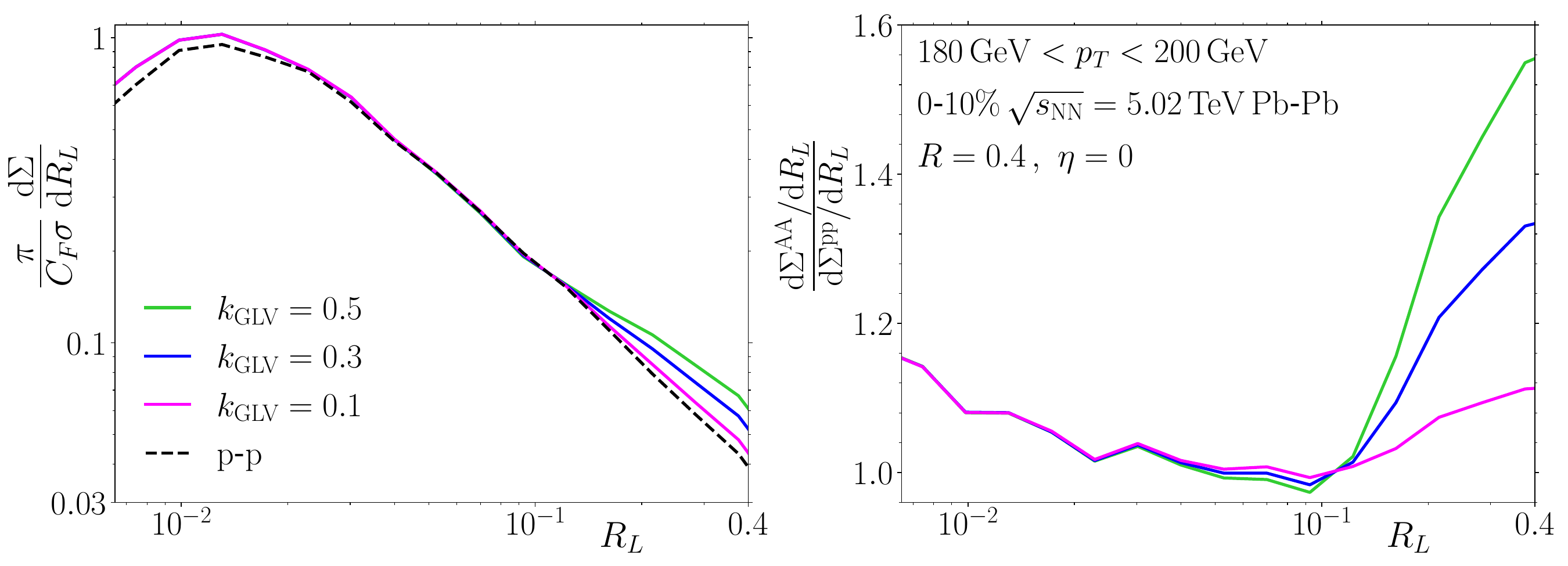}
\caption{Left panel: EEC for inclusive $180 <p_T <200$\,GeV  jets within the 0-10$\%$ centrality class in $\sqrt{s_{\rm NN}} =5.02$ TeV Pb-Pb collisions computed through eq.~\eqref{eq:eecinclusivjet} using the single scattering approach described in  \ref{subsec:GLV} for several values of the free parameter $k_{\rm GLV}$ (solid curves) compared to the p-p NLL EEC (dashed black curve). Right panel: Ratio of the Pb-Pb EEC w.r.t the p-p EEC.  }
\label{fig:GLV3.2_180}
\end{figure}

\bibliographystyle{JHEP}

\bibliography{qgp_EEC_ref.bib}

\end{document}